\documentstyle[sprocl,epsfig]{article}
\def\bge{\begin{equation}}
\def\ene{\end{equation}}
\def\bg{\begin{eqnarray}}
\def\en{\end{eqnarray}}

\def\bge{\begin{equation}}
\def\ene{\end{equation}}
\def\bg{\begin{eqnarray}}
\def\en{\end{eqnarray}}

\def\ra{\rightarrow}

\def\What{\widehat{W}}
\def\Wzero{\widehat{W}_0}
\def\Wone{\widehat{W}_1}
\def\Wtwo{\widehat{W}_2}

\def\pslash{\not\!p}
\def\qslash{\not\!q}

\def\S0{{\Sigma^0}}

\def\X0{{\Xi^0}}

\newcommand{\AmS}{{\protect\the\textfont2
  A\kern-.1667em\lower.5ex\hbox{M}\kern-.125emS}}

\begin{document}

\title{NON-PERTURBATIVE STRUCTURE OF THE NUCLEON}

\author{A. W. THOMAS}

\address{Department of Physics and Mathematical Physics, \\
and Special Research Centre for the Subatomic Structure of Matter, \\
University of Adelaide,\\ Adelaide, 5005 Australia \\
E-mail: athomas@physics.adelaide.edu.au}


\maketitle\abstracts{While much attention has been focussed on the
successes of perturbative QCD 
in describing the $Q^2$-dependence of deep-inelastic 
structure functions, the starting distributions 
themselves contain important,
non-perturbative information on the structure of the nucleon, which has
been somewhat neglected. We review
some of the most important, recent discoveries resulting from studies of
deep-inelastic scattering. There are important connections between these
discoveries and low energy properties of the nucleon and wherever
possible we shall make these clear. In particular, we shall see that
well known features of QCD, such as dynamical symmetry breaking, are
reflected in the properties of the measured parton distributions.}

\section{Introduction}

There have recently been some very promising advances in lattice QCD as
a result of the development of improved actions\cite{LL}. 
In addition, we have a
wealth of sophisticated models of hadron structure built on our
knowledge of QCD, especially its symmetries. While lattice QCD cannot
yet compete with such models in certain critical areas of hadron
structure, such as dynamical symmetry breaking, all theoretical
approaches can benefit from precise experimental insights into the
problem. Apart from the usual low energy properties, such as masses,
charge radii and magnetic moments, there is a wealth of information
available from deep-inelastic scattering which is just beginning to be
taken seriously as a source of information on 
non-perturbative physics\cite{DISS}.

We shall review the latest experimental information on the light-quark
sea of the nucleon, which shows a dramatic deviation from the naive
expectations of perturbative QCD. From this starting point it is natural
to ask about the strange and even the charm components of the sea of the
nucleon. These components will be the focus of considerable
investigation in the near future. From the sea we turn to the behaviour
of the valence quark distributions and particularly the behaviour of the
distributions at large-$x$. Here too there are a number of surprises
from the most recent analysis of the experimental data and, in the case
of spin-dependent distributions, some interesting ideas to be tested.

\section{SU(2)$_F$ Violation}

The fundamental degrees of freedom in QCD are quarks and gluons, and for
a considerable time there was great reluctance to include any other
degrees of freedom in modeling hadron structure.
On the other hand, extensive studies of non-perturbative QCD have shown 
that the chiral symmetry of the QCD Lagrangian is dynamically broken and 
that the resulting, massive constituent quarks must be coupled to pions.
As pseudo-Goldstone bosons, the latter would be massless in the chiral 
limit.
Most importantly, as emphasised by extensive work on chiral perturbation 
theory, {\em no perturbative treatment of $q \bar{q}$ creation and 
annihilation can ever generate the non-analytic behaviour in the light
quark mass for physical quantities (such as $M, <r^2>, \sigma_{\pi N}$) 
which are generated by these Goldstone bosons}.
As a consequence, it is now difficult to imagine a realistic quark model
which does not incorporate at least the pion cloud of the nucleon.
\begin{figure}[htb]
\begin{center}
\epsfig{file=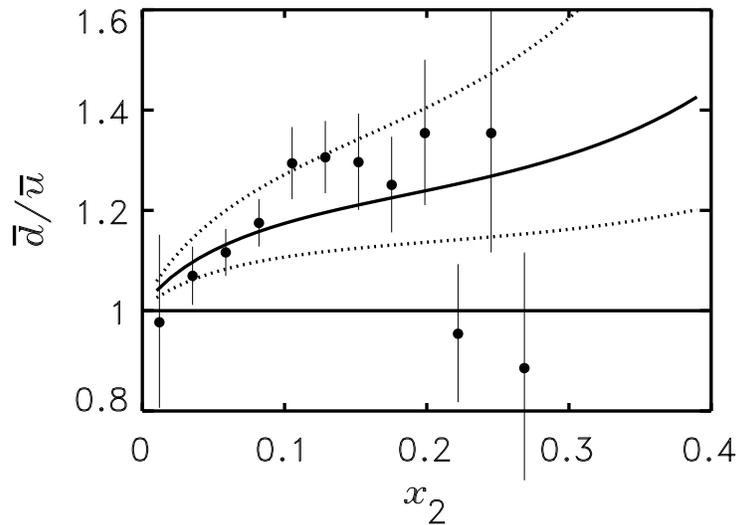,height=8cm}
\caption{Comparison of the value of $\bar{d}/\bar{u}$ extracted from the
	recent E866 data \protect\cite{E866} with the expectations from
	just the $\pi N$ Fock component of the nucleon wave function,
	for three different form factors at the $\pi N$ vertex -- from
	Ref. \protect\cite{MT_fewb}.}
\end{center}
\end{figure}

The development of chiral quark models began in the late 1970's
\cite{IM,CHT,BRR,CBM} and it became clear that the pion field had very
important practical consequences for the low energy properties of hadrons.
For example, the coupling to a ``bare'' nucleon lowers its mass by several
hundred MeV, and can contribute a large part of the $N-\Delta$ mass 
difference, while the charge form factor of the neutron is a first
order effect of its pion cloud \cite{CBM}. 

However, the relevance of the pion cloud for deep inelastic scattering,
which had been first realized by Sullivan \cite{SULL},
was not explored in depth until the discovery of the nuclear EMC effect,
when it was suggested that the effect might be caused by a nuclear
enhancement of the pion field of the bound nucleons \cite{CLLS,ET}.
At this time it was realized that the light mass of the pion would lead
to an enhancement of the non-strange over the strange sea of the nucleon
\cite{THO83}, and this was used to put a constraint on the pion--nucleon
form factor --- a limit that has since been explored in detail \cite{FMS}.
A more important consequence of the nucleon's pion cloud, which was also 
pointed out in Ref.\cite{THO83}, was an excess of $\bar{d}$ over $\bar{u}$
quarks.
In particular, simple Clebsch-Gordan coefficients for isospin show that
the pion cloud of the proton is in the ratio 2:1 for $\pi^+:\pi^0$.
Since the $\pi^+$ contains only a valence $\bar{d}$ and the $\pi^0$
equal amounts of $\bar{d}$ and $\bar{u}$, this component of the pion
cloud of the nucleon yields a ratio for $\bar{d}:\bar{u}$ of 5:1.

On the other hand, perturbative QCD inevitably leads to the conclusion
that $\bar{u} = \bar{d}$, a result known as SU(2)-flavor, SU(2)$_F$, 
symmetry.
The latter is a very misleading name as there is {\em no rigorous
symmetry involved}.
Indeed, the prediction that the component of the sea arising from
the long-range piece of the pion cloud of the nucleon satisfies 
$\bar{d}:\bar{u}$ = 5:1 is totally consistent with charge independence.
Thus the violation of the Gottfried sum rule and the subsequent
measurement of $\bar{d} - \bar{u}$ gives us direct evidence that there
is a sizable non-perturbative component of the nucleon sea.
Various studies of the pion cloud of the nucleon since the original NMC
measurement \cite{NMCO} have concluded that this is indeed the most
likely explanation of the observed violation of the sum rule 
\cite{SPTH,HM,SST,KL,HSB}.

In Fig.1 we show the calculated ratio of $\bar{d}/\bar{u}$ from the
$\pi N$ Fock component of the nucleon wave function as a source of 
asymmetry, in comparison with the data points extracted from the 
preliminary results \cite{E866} from the E866 collaboration on the
ratio of $pD$ and $pp$ Drell-Yan cross sections.
The calculation was performed in the light-cone formalism \cite{MTLC,ZOLL}
with a monopole $\pi NN$ form factor mass parameter $\Lambda = 0.7,
1.0$ and 1.3 GeV --- from smallest to largest.
Clearly the agreement is qualitatively excellent, but a more detailed
analysis needs to be carried out once all the data have been analysed.
Let us emphasise once more the importance of these data, which are giving
us direct insight into the way dynamical chiral symmetry breaking is
realized in the nucleon.

\section{Intrinsic strangeness and charm}

The experience with the breaking of SU(2)$_F$ symmetry, which we have
just described, leads us to take much more seriously the possibility of
an intrinsic component of the strange and charm quark sea.
The former was first discussed by Signal and Thomas \cite{ST} and has
recently been investigated by a number of authors \cite{BROD,SMSBAR}
in the light of new neutrino data from CCFR \cite{CCFR}. The implication
of a non-perturbatively generated component of the strange sea is a
marked asymmetry between $s$ and $\bar{s}$, which 
will clearly be the subject of much more detailed investigation in 
future.

With regard to the question of intrinsic charm, there has been some
suggestion that it may play a role in the anomalous events seen recently
at HERA -- see section 5 below. For example, Gunion and Vogt \cite{GV}
examined a model of the 5-quark component of the nucleon 
wave function on the light-cone \cite{BHPS}.
{}Following Brodsky {\em et al.} \cite{BHPS}, the wave function was 
assumed to be inversely proportional to the light-cone energy difference
between the nucleon ground state and the 5-quark excited state.
The resulting $x$-dependence of the inclusive $c$ quark distribution
in the minimal model of \cite{GV} was given by \cite{BHPS}:
\begin{eqnarray}
\label{delic}
\delta^{(IC)} c(x)
&=& 6 x^2 \left( (1-x) (1 + 10x + x^2) - 6x (1+x) \log 1/x \right),
\end{eqnarray}
with the normalization fixed to 1\%.
Such a distribution peaks at $x \sim 0.2$, and is
negligible beyond $x \sim 0.7$.
The anti-charm distribution is assumed to be equal to the charm
distribution in this model,
$\delta^{(IC)} \overline c(x) = \delta^{(IC)} c(x)$.

As an alternative to the intrinsic charm picture of Refs.\cite{GV,BHPS},
in Refs.\cite{NNNT,PNNDB} the charmed sea was taken to arise from the
quantum fluctuation of the nucleon to a virtual $D \Lambda_c$
configuration -- by analogy with the successful description of the
observed $\bar{d} - \bar{u}$ asymmetry in the light-quark sector.
The nucleon charm radius \cite{NNNT} and the charm quark distribution 
\cite{PNNDB} were both estimated in this framework.
Whether the same philosophy can be justified for a cloud of heavy charmed
mesons and baryons around the nucleon is rather more questionable given
the large mass of the fluctuation.
Nevertheless, to a first approximation, we may take the meson cloud
framework as an indicator of the possible shape
of the non-perturbative charm distribution.
A natural prediction of this model are highly asymmetric $c$
and $\overline c$ distributions.

In the meson cloud model, the distribution of charm quarks in the
nucleon on the light cone at some low hadronic scale is written
in convolution form\cite{MT_HERA}:
\begin{eqnarray}
\delta \overline c(x)
&=& \int_x^1 {dz \over z} f_{D/N}(z)\ 
	\overline c^{D}\left({x \over z}\right),\ \ 
\delta c(x)
\ =\ \int_x^1 {dz \over z} f_{\Lambda_c/N}(z)\ 
	c^{\Lambda_c}\left({x \over z}\right),
\end{eqnarray}
where $z$ is the fraction of the nucleon's light-cone momentum 
carried by the $D$ meson or $\Lambda_c$.
The light cone (or infinite momentum frame) distribution of $D$
mesons in the nucleon is given by:
\begin{equation}
\label{fz}
f_{D/N}(z)
=\int_0^\infty \frac{dk^2_\perp}{16\pi^2}
{ g^2(k_\perp^2,z) \over z (1-z) (s_{D \Lambda_c} - M_N^2)^2 }
\left( { k_\perp^2 + [M_{\Lambda_c} - (1-z) M_N]^2 \over 1-z } \right),
\end{equation}
and can be shown to be related to the light-cone distribution of 
$\Lambda_c$ baryons, $f_{\Lambda_c/N}(z)$,
by $f_{\Lambda_c/N}(z) = f_{D/N}(1-z)$.
In Eq.(\ref{fz}) the function $g$ describes the extended nature of the
$D\Lambda_c N$ vertex, with the momentum dependence parameterized by\
$g^2(k_\perp^2,z)\
=\ g_0^2\ (\Lambda^2 + M_N^2)/(\Lambda^2 + s_{D\Lambda_c})$,
where the $D\Lambda_c$ center of mass energy squared
is given by\ 
$s_{D \Lambda_c} = (k_\perp^2 + M_D^2)/z
		 + (k_\perp^2 + M_{\Lambda_c}^2)/(1-z)$,
and $g_0$ is the $D \Lambda_c N$ coupling constant at the
pole, $s_{D \Lambda_c} = M_N^2$.
We expect $g_0$ to be similar to the $\pi NN$ coupling constant.

Because of the large mass of the $c$ quark, one can approximate the
$\overline c$ distribution in the $D$ meson \cite{PNNDB} and the
$c$ distribution in the $\Lambda_c$ by:
\begin{eqnarray}
\overline c^{D}(x) &\approx& \delta(x-1),\ \ \
c^{\Lambda_c}(x)\ \approx\ \delta(x-2/3),
\end{eqnarray}
which then gives:
\begin{eqnarray}
\label{mcm_final}
\delta \overline c(x) &\approx& f_{D/N}(x),\ \ \ \ \ \ 
\delta c(x)\ \approx\ {3 \over 2} f_{\Lambda_c/N}(3x/2).
\end{eqnarray}
\begin{figure}[hbt]
\begin{center}
\epsfig{file=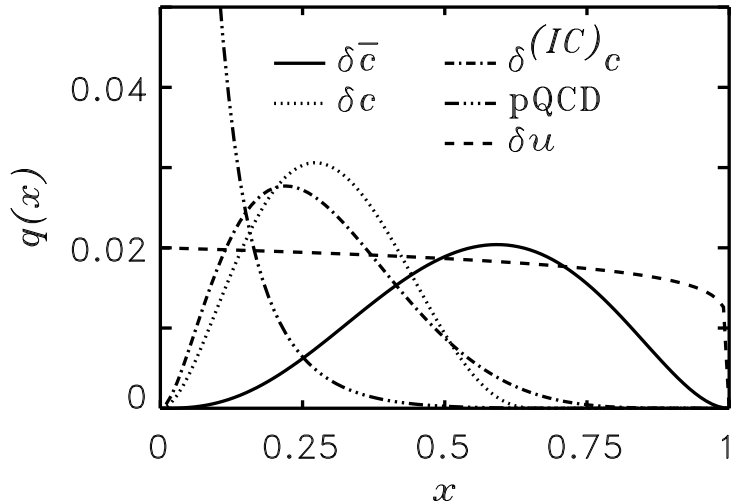,height=8cm}
\caption{The non-perturbative $\delta c$ and $\delta \overline c$
distributions in the meson cloud model \protect\cite{MT_HERA} (the latter
normalized to 1\%). Also shown is the light-cone intrinsic charm
distribution, $\delta^{(IC)} c$, of Refs.\protect\cite{GV,BHPS} and the
modification $\delta u$ of the valence $u$ distribution from 
Ref.\protect\cite{KLT}. The purely perturbative contribution at
$Q^2=$4~GeV$^2$ is also shown.}
\end{center}
\end{figure}
The resulting $\delta c$ and $\delta \overline c$ distributions are shown
in Fig.2, calculated for an ultraviolet cutoff of $\Lambda \approx 2.2$~GeV,
which gives
$\int_0^1 dx \delta c(x) = \int_0^1 dx \delta \overline c(x) \approx 1\%$.
For a probability of 0.5\% one would need a smaller cutoff,
$\Lambda \approx 1.7$~GeV.
Quite interestingly, the shape of the $c$ quark distributions is  
similar to that in the intrinsic charm model of Refs.\cite{GV,BHPS}.
However, as mentioned above, the model of \cite{GV,BHPS} assumes identical
shapes for the non-perturbative $c$ and $\overline c$ distributions, while
the meson cloud gives a significantly harder $\overline c$ distribution.
It is a very important issue for our understanding of strong interaction
dynamics whether or not the charm quark distributions exhibit such an
asymmetry. 

\section{The role of perturbative QCD at large-$x$}

The precise mechanism whereby the SU(6), spin-flavor symmetry of the
parton
distributions of the nucleon is broken is a question of fundamental
importance in hadronic physics.
The SU(6) spin-flavour wave function of a proton,
polarized in the $+z$ direction, has the form \cite{CLO79}:
\begin{eqnarray}
\label{pwfn}
\left| p\uparrow \rangle \right.
&=& {1 \over \sqrt{2}} \left| u\uparrow (ud)_{S=0} \rangle \right. \
 +\ {1 \over \sqrt{18}} \left| u\uparrow (ud)_{S=1} \rangle \right. \
 -\ {1 \over 3}  \left| u\downarrow (ud)_{S=1} \rangle \right. \
\nonumber \\
& &
 -\ {1 \over 3}  \left| d\uparrow (uu)_{S=1} \rangle \right. \
 -\ {\sqrt{2} \over 3} \left| d\downarrow (uu)_{S=1} \rangle \right.,
\end{eqnarray}
For the neutron to proton structure function ratio this implies:
\begin{eqnarray}
{ F_2^n \over F_2^p }
&=& {2 \over 3}\ \ \ \ \ \  ; \ \ \ \ {\rm SU(6)\ symmetry}.
\end{eqnarray}

Of course, SU(6) spin-flavor symmetry is not exact.
The nucleon and $\Delta$ masses are split by some 300 MeV and
empirically
the $d$ quark distribution is softer than the $u$.
The correlation between the mass splitting in the {\bf 56} baryons and
the
large-$x$ behavior of $F_2^n/F_2^p$ was observed some time ago by Close
\cite{CLO73} and Carlitz \cite{CAR75}.
Based on phenomenological arguments,
the breaking of the symmetry in Eq.(\ref{pwfn}) was argued to arise from
a suppression of the ``diquark'' configurations having $S=1$ relative to
the $S=0$ configuration.
Such a suppression is, in fact, quite natural if one observes that
whatever
mechanism leads to the observed $N-\Delta$ splitting (e.g.
color-magnetic
force, instanton-induced interaction, pion exchange), necessarily acts
to
produce a mass splitting between the possible spin states of the
spectator
pair, $(qq)_S$, with the $S=1$ state heavier than the $S=0$ state by
some
200 MeV \cite{CT}.
{}From Eq.(\ref{pwfn}), a dominant scalar valence diquark component of
the
proton suggests that in the $x \rightarrow 1$ limit $F_2^p$ is
essentially
given by a single quark distribution (i.e. the $u$), in which case:
\begin{eqnarray}
{ F_2^n \over F_2^p }
&\rightarrow& { 1 \over 4 }, \ \ \ \ \
{ d \over u } \rightarrow 0\ \ \ \ \
; \ \ \ \ S=0\ {\rm dominance}.
\end{eqnarray}
This expectation has, in fact, been built into all phenomenological fits
to the parton distribution data.

An alternative suggestion, based on perturbative QCD, was originally
formulated by Farrar and Jackson \cite{FJ}.
There it was argued that the exchange of longitudinal gluons, which are
the only type permitted when the spin projections of the two quarks in
$(qq)_S$ are aligned, would introduce a factor $(1-x)^{1/2}$ into the
Compton amplitude --- in comparison with the exchange of a transverse
gluon between quarks with spins anti-aligned.
In this approach the relevant component of the proton valence wave
function
at large $x$ is that associated with states in which the total
``diquark''
spin {\em projection}, $S_z$, is zero.
Consequently, scattering from a quark polarized in the opposite
direction
to the proton polarization is suppressed by a factor $(1-x)$ relative to
the helicity-aligned configuration.
\begin{figure}[hbt]
\begin{center}
\epsfig{file=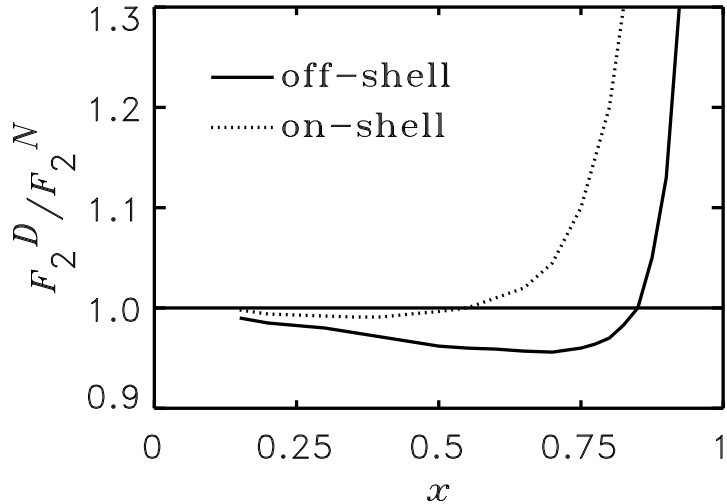,height=8cm}
\caption{$F_2^D/F_2^N$ ratio as a function of $x$ for the model of
	Refs.\protect\cite{MST1,MST2} (solid) which accounts for
	off-shell kinematics, and the on-shell model of
Ref.\protect\cite{FS} (dotted) --- from
Ref.\protect\cite{MST2}.}
\end{center}
\end{figure}

A similar result is also obtained in the treatment of Brodsky {\em et
al.}
\cite{BBS} (based on counting-rules), where the large-$x$ behavior of
the
parton distribution for a quark polarized parallel ($\Delta S_z = 1$) or
antiparallel ($\Delta S_z = 0$) to the proton helicity is given by:
$q^{\uparrow\downarrow}(x) = (1~-~x)^{2n - 1 + \Delta S_z}$,
where $n$ is the minimum number of non-interacting quarks (equal to 2
for
the valence quark distributions).
In the $x \rightarrow 1$ limit these arguments, based on PQCD, suggest:
\begin{eqnarray}
{ F_2^n \over F_2^p }
&\rightarrow& {3 \over 7}, \ \ \ \ \
{ d \over u } \rightarrow { 1 \over 5 }\ \ \ \ \
; \ \ \ \ S_z=0\ {\rm dominance}.
\end{eqnarray}
Note that the $d/u$ ratio {\em does not vanish} in this case.
Clearly, if one is to understand the dynamics of the nucleon's quark
distributions at large $x$, it is imperative that the consequences of
these models be tested experimentally.

\section{Reanalysis of the experimental data at large-$x$}

Information on the structure functions of the neutron is obtained from
the analysis of deep-inelastic scattering data on the deuteron
\cite{WHIT,GOMEZ}.
Amongst the many approaches to this problem we mention the light-front
treatment \cite{FS,KU} and the relativistic impulse approximation
\cite{LG,BT}, involving the free nucleon structure function at a shifted
value of $x$ or $Q^2$ \cite{DT,NW}.
A more phenomenological approach, developed by Frankfurt and Strikman
\cite{FS88}, attempts to derive the nuclear correction in the deuteron
by extrapolation from higher $A$ as a function of the ``effective
density''
of the nucleus.
The experimental extraction of $F_2^n$ is usually made using either the
phenomenological effective density approach or the older, ``pre-EMC''
theoretical treatments.

Within traditional nuclear physics the natural approach to the structure
function of the deuteron is the impulse approximation.
This assumes a convolution of the free nucleon
structure function, $F_2^N$, with the non-relativistic momentum
distribution $f_{N/D}$ of nucleons in the deuteron, calculated
in terms of a non-relativistic wave function of the deuteron and
its binding energy.
Although the binding energy is very small, the kinetic energy of the
recoiling nucleon plays a significant role in forcing the struck nucleon
further off-shell than one would usually expect\cite{DT}.

In order to assess the theoretical reliability of the non-relativistic
impulse approximation one needs to go beyond the usual assumptions made
in the convolution approach \cite{CONV}.
In particular, the ingredients necessary for a covariant, relativistic
description are a covariant $DNN$ vertex with one of the nucleons
(the spectator to the hard collision) on-mass-shell, and an off-shell
photon--nucleon scattering amplitude (``off-shell nucleon structure
function'') $\widehat W$, the full structure of which was only recently
derived in Ref.\cite{MST1}.

The analysis of Ref.\cite{MST1} showed that the most general form of
the operator $\What$ (which is a $4\times 4$ matrix in Dirac space),
consistent with the discrete symmetries and gauge invariance, which
contributes in the Bjorken limit is:
\begin{equation}
\What = \Wzero I + \Wone \pslash + \Wtwo \qslash,
\label{3}
\end{equation}
where the $\widehat W_i$ are functions of $p^2, q^2$ and $p \cdot q$
($q$ is the virtual photon four-momentum).
Thus, whereas in the free case the nucleon structure function involves
the combination:
\begin{equation}
{\rm Tr}[(\pslash + M) \What] \sim M \Wzero + M^2 \Wone + p \cdot q
\Wtwo,
\label{4}
\end{equation}
the deuteron structure function involves:
\begin{equation}
{\rm Tr}[(A_0 + \gamma^\mu A_{1 \mu}) \What]
\sim A_0 \Wzero + p \cdot A_1 \Wone + q \cdot A_1 \Wtwo.
\label{5}
\end{equation}
Clearly then, even in the absence of Fermi motion, one finds that
in general $F_2^D \neq F_2^N$.

Having established that, in principle, the structure function of the
bound
nucleon cannot equal the structure function of the free nucleon, the
important question is how big the difference actually is in practice.
To estimate this, one can construct a simple model \cite{MST1} of the
so-called hand-bag diagram for an off-shell nucleon, in which the
$N$-quark
vertex is taken to be either a simple scalar or pseudo-vector, with the
parameters adjusted to reproduce the free nucleon structure functions
\cite{MSM}.

The result of the fully off-shell calculation from Ref.\cite{MST2}
is shown in Fig.3 (solid curve), where the ratio of the total
deuteron to nucleon structure functions ($F_2^D/F_2^N$) is plotted.
(We note that the behaviour of the full off-shell curve in Fig3.
is qualitatively similar to that found by Uchiyama and Saito \cite{US},
Kaptari and Umnikov\cite{KU}, and Braun and Tokarev \cite{BT}.)
We also show the result of an on-mass-shell calculation from
Ref.\cite{FS} (dotted curve), which has been used in many previous
analyses of the deuteron data \cite{EMC,WHIT}.
The most striking difference between the curves is the fact that the
on-shell ratio has a very much smaller trough at $x \approx 0.3$, and
rises faster above unity (at $x \approx 0.5$) than the off-shell curve,
which has a deeper trough, at $x \approx 0.6-0.7$, and rises above unity
somewhat later (at $x \approx 0.8$).

Clearly, a smaller $D/N$ ratio at large $x$, as in
Refs.\cite{MST1,MST2},
implies a larger neutron structure function in this region.
To estimate the size of the effect on the $n/p$ ratio requires one to
extract $F_2^n$, while taking care to eliminate any effects  
that may arise from the extraction method itself.
Melnitchouk and Thomas \cite{MTNP} therefore used exactly the same
extraction procedure as used in previous EMC \cite{EMC} and SLAC
\cite{WHIT} data analyses, namely the smearing (or deconvolution)
method discussed by Bodek {\em et al.} \cite{BODEK}.
{\em This method involves the direct use of the proton and deuteron
data, without making any assumption concerning $F_2^n$ itself}.

The results of this analysis are presented in Fig.4, using both the
off-shell calculation \cite{MST2} (solid points) and the on-shell model
\cite{FS} (open points).
The increase in the ratio at large $x$ for the off-shell case 
is a direct consequence
of the deeper trough in the $F_2^D/F_2^N$ ratio in Fig.3.
We notice, in particular, that the values of $F_2^n/F_2^p$ 
obtained with the off-shell
method appear to approach a value broadly consistent with the
Farrar-Jackson \cite{FJ} prediction of 3/7, whereas the data previously
analyzed in terms of the on-shell formalism produced a ratio tending to
the lower value of 1/4.
\begin{figure}[hbt]
\begin{center}
\epsfig{file=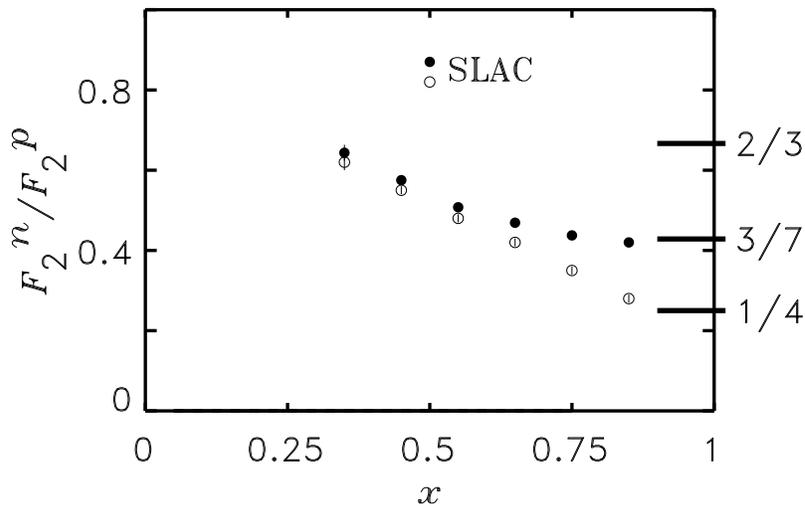,height=8cm}
\caption{Deconvoluted $F_2^n/F_2^p$ ratio extracted from the
	SLAC $p$ and $D$ data \protect\cite{WHIT,GOMEZ} using
		the model of Ref.\protect\cite{MST1,MST2} (solid circles)
		and Ref.\protect\cite{FS} (open circles) -- from 
		Ref. \protect\cite{MTNP}.}
\end{center}
\end{figure}

The $d/u$ ratio, shown in Fig.5, is obtained by inverting $F_2^n/F_2^p$
in the valence quark dominated region.
The points extracted using the off-shell formalism (solid circles) are
again significantly above those obtained previously with the aid of the
on-shell prescription.
In particular, they indicate that the $d/u$ ratio may actually approach
a {\em finite} value in the $x \rightarrow 1$ limit, contrary to the
expectation of the model of Refs.\cite{CLO73,CAR75}, in which $d/u$
tends to zero.
Although it is {\em a priori} not clear at which scale these model
predictions should be valid, for the values of $Q^2$ corresponding
to the analyzed data the effects of $Q^2$ evolution are minimal.
\begin{figure}[hbt]
\begin{center}
\epsfig{file=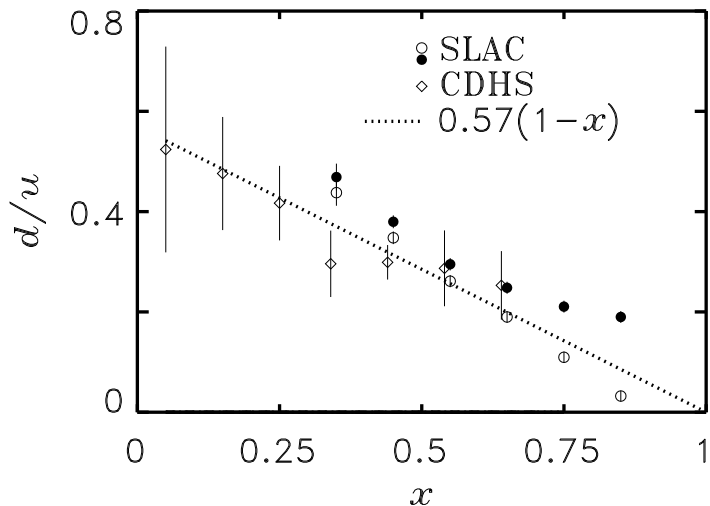,height=8cm}
\caption{The $d/u$ ratio extracted from the results of Fig.4.
	Also shown for comparison is the ratio extracted from neutrino
		measurements by the CDHS collaboration \protect\cite{CDHS} and
		(dotted line) a standard linear fit.}
\end{center}
\end{figure}

Naturally, it would be preferable to extract $F_2^n$ at large $x$
without having to deal with uncertainties in the nuclear effects.
In principle this could be achieved by using neutrino and
antineutrino beams to measure the $u$ and $d$ distributions in
the proton separately, and reconstructing $F_2^n$ from these.
Unfortunately, as seen in Fig.5, the neutrino data 
do not extend out to very large
$x$ ($x > 0.6$), and at present cannot discriminate between
the different methods of analyzing the electron--deuteron data.

\section{The HERA Anomaly}

The H1 and ZEUS experiments at HERA have recently produced a small
number of events at enormously high $Q^2$ which have generated
tremendous theoretical interest \cite{H1,ZEUS}.
For $Q^2 > 10,000 $GeV$^2$ and $x > 0.45$ the valence parton
distributions
are calculated to drop dramatically.
The HERA anomaly is essentially the excess of observed ``neutral
current'' (NC) 
events (i.e., events of the type $e^+ p \ra e^+ X$)
over expectations by roughly
an order of magnitude.
Many exotic explanations of this excess have already been suggested,
indeed,
the number of possibilities currently exceeds the number of events.
However, before the new physics can be worked out one must be sure that
the
input parton distributions used to estimate ``background'' rates are
reliable.

One glaring problem with the current treatment of the partonic
``background'' is that {\em all} of the standard distributions used are
constructed to satisfy $d/u \ra 0$ as $x \ra 1$ at low-$Q^2$.
As we have seen, the recent re-analysis of the deuteron data leads to a
$d/u$ ratio which appears to be consistent with the prediction of PQCD
that $d/u \ra 1/5$ as $x \ra 1$.
In the light of this result it is not only vital to find alternative,
more direct measurements of $d/u$ at large $x$, but those generating
standard sets of parton distributions should at the very least present
alternative parameter sets consistent with the new analysis of the
deuteron data.
Until parameter sets are constructed which are consistent with
$d/u \ra 1/5$ as $x \ra 1$, at $Q^2 \sim 10$ GeV$^2$, one cannot be
sure of the reliability of ``background'' rate estimates at the extreme
values of $Q^2$ and $x$ being probed at HERA.

\begin{figure}[htb]
\begin{center}
\epsfig{file=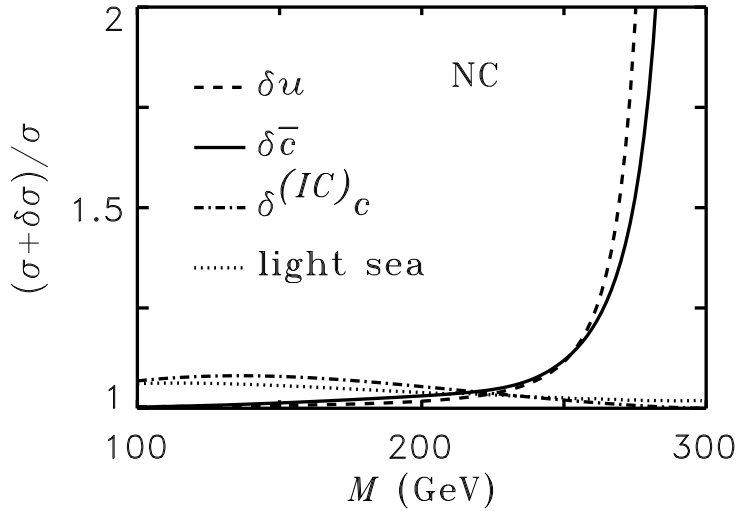,height=8cm}
\caption{Ratio of modified to standard DIS model NC cross sections
	at the charm threshold (from Ref. \protect\cite{MT_HERA}), 
	with the modifications arising from
	the additional $u$ quark component \protect\cite{KLT} (dashed),
	1\% non-perturbative $\delta \overline c$ and $\delta c$
	distributions from the meson cloud model (solid),
	and the intrinsic charm model of
	Refs.\protect\cite{GV,BHPS} (dot-dashed). Also shown is the effect
	of the meson cloud contributions to the light sea quarks (dotted).}
\end{center}
\end{figure}
The effect of an intrinsic charm component of the sea of the nucleon was
recently examined by Gunion and Vogt \cite{GV}, with the conclusion that,
as calculated on the basis of counting rules, it was too soft to explain
the observed anomaly. What one needs, therefore,
is a substantially harder distribution which
has significantly more strength above $x \sim 0.6$ than in Eq.(\ref{delic}).
This was precisely what was found in the non-perturbative calculation of
Melnitchouk and Thomas \cite{MT_HERA}  -- see Fig.2 above.

The calculation of the NC and CC cross sections requires parton distributions
for all flavors.
For this we use a recent parameterization of global data from the CTEQ
Collaboration \cite{CTEQ}.
Expressions for the differential NC and CC cross sections $d^2\sigma/dxdQ^2$
in the standard model can be found in Refs.\cite{ZEUS} and \cite{H196}.
In Fig.6 we show the ratios of the modified to standard DIS model NC
cross sections, with $\sigma \equiv d^2\sigma/dxdQ^2$, and 
$\sigma + \delta\sigma$
represents the cross section calculated with the modified distributions.
The result with the modified $u$ distribution \cite{KLT}, which was
rather artificially created to reproduce the HERA anomaly, rises sharply
above $M \sim 200$ GeV.
The effect is rather similar if one includes the non-perturbative $\delta c$
and $\delta \overline c$ distributions from the meson cloud model.
On the other hand, with the somewhat softer, intrinsic charm distribution
of Refs. \cite{GV,BHPS}, the enhancement is rather modest, and less than about
10\% over the whole range of $M$.
\begin{figure}[htb]
\begin{center}
\epsfig{file=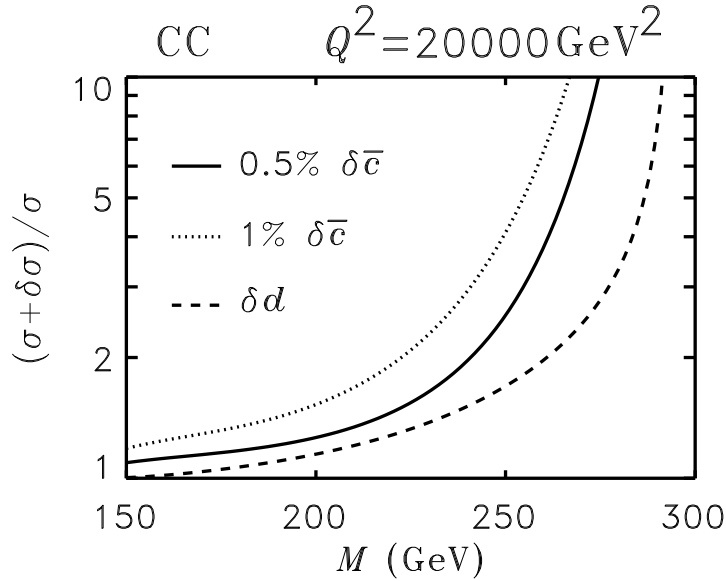,height=8cm}
\caption{Ratio of modified to standard DIS model CC cross sections at
$Q^2=20000$~GeV$^2$, with the modifications arising from
0.5\% and 1\% additional $\delta \overline c$ distributions,
as well as a modified $d$ quark distribution at large $x$
-- from Ref. \protect\cite{MT_HERA}.}
\end{center}
\end{figure}

We also show the effect of additional contributions to the light
flavor distributions ($u, \overline u, d$ and $\overline d$) which one would
obtain from the pion cloud of the nucleon.
Although these are considerably larger in magnitude than the $\delta c$
or $\delta \overline c$ distributions, because they appear at small $x$
($\sim 0.1$) their effect is to yield only a very small enhancement of 
the cross section ratio.
{}From this figure one can conclude that the only realistic candidates
for a significant enhancement of the cross section at large $M$ are the
modified valence $u$ distribution from Ref.\cite{KLT}, and the hard charm
distributions in Eq.(\ref{mcm_final}).

For scattering via the CC, the effect of the additional contributions to
the parton distributions is shown in Fig.7 for $Q^2 = 20000$~GeV$^2$.
Since the $W^+$ boson is not sensitive to the $u$ quark in the proton,
the $\delta u$ modification has no effect on the $e^+ p$ cross section.
In contrast, as noted by Babu {\it et al.} \cite{BABU}, the effect of
an additional non-perturbative $\delta \overline c$ contribution is an
even larger enhancement of the CC cross section than the NC cross section.
With a 0.5\% (1\%) intrinsic charm component the CC cross section
increases by a factor $\sim$ 2 (3) for $200 < M < 250$~GeV,
which is similar to the excess observed by H1 \cite{H1} in this region.

\section{Spin Dependent Structure Functions}

The spin structure functions of the nucleon, $g_1^{p(n)}$, are of
tremendous interest at present.
Experimentally, $g_1$ is proportional to the difference of DIS cross
sections for $ep$ scattering with beam and target helicities aligned
and anti-aligned \cite{CHENG}. 
Within the parton model it may be written in terms of the parton helicity 
(loosely ``spin'') distributions,
$\Delta q(x) = [q^{\uparrow} - q^{\downarrow}
	       + \bar{q}^{\uparrow} - \bar{q}^{\downarrow}]$,
with $q^{\uparrow (\downarrow)}$ the number density of quarks with
helicity parallel (anti-parallel) to the helicity of the target proton:
\bge
g_1^p(x) = \frac{1}{2} \sum_q e^2_q \Delta q(x).
\label{eq:3.1}
\ene

Intense interest in the spin structure functions began in 1988 when EMC
announced a large violation of the Ellis-Jaffe sum rule \cite{EMC_SPIN}, 
which relates $\Gamma_p(Q^2) \equiv \int^1_0 g_1^p(x,Q^2) dx$ to the
isovector and octet axial-vector coupling constants, $g_A^{(3)}$ and
$g_A^{(8)}$.
The failure of this sum rule, which is not a rigorous consequence of QCD,
led to questions about the Bjorken sum rule, which relates
$\Gamma_p - \Gamma_n$ to $g_A^{(3)}/6$ (modulo QCD radiative corrections
\cite{VER}) and {\em is} a strict consequence of QCD.
To determine $\Gamma_n$ one must measure $g_1^n(x)$, which requires a
polarized nuclear target such as $^3$He or D.
At present, all neutron data extracted from the deuteron are obtained
by applying a simple, non-relativistic prescription to correct $g_1^D$ 
for the $D$-state component (probability $\omega_D$) of the deuteron
wave function \cite{DPOL}:
\bge
g_1^n(x)
= \left(1 - \frac{3}{2} \omega_D\right)^{-1} g_1^D(x) - g_1^p(x).
\label{eq:3.3}
\ene
As we explain below, exactly the same techniques described in 
section 5 may be used to test the accuracy of Eq.(\ref{eq:3.3}).

While most interest has been focussed on the issue of sum rules,
we stress that the shapes of $g_1^p(x)$ and $g_1^n(x)$ contain even
more important information.
For example, the same arguments that led to different conclusions
about the behaviour of $d/u$ as $x \ra 1$ also give quite different 
predictions for $g_1^p$ and $g_1^n$ as $x \rightarrow 1$, namely 1/4
and 3/7, respectively.
Quite interestingly, while the ratio of the 
polarized to unpolarized $u$
quark distributions is predicted to be the same in the two models:
\begin{eqnarray}
\label{Deltau}
{ \Delta u \over u }
&\rightarrow& 1\ \ \ \ ; \ \ \ \ S=0\ or\ S_z=0\ {\rm dominance},
\end{eqnarray}
the results for the $d$-quark distribution ratio differ even in sign:
\begin{eqnarray}
{ \Delta d \over d }
&\rightarrow& - {1 \over 3}\ \ \ \ ; \ \ \ \ S=0\ {\rm dominance},\\
&\rightarrow& 1\ \ \ \ ; \ \ \ \ S_z=0\ {\rm dominance}.
\label{Deltad}
\end{eqnarray}

Using the same techniques described earlier for the
unpolarized case, Melnitchouk, Piller and Thomas \cite{MPT} derived the
most general, antisymmetric, Dirac tensor operator of twist 2 for an
off-mass-shell nucleon (see also \cite{KMPW}):
\bge
\widehat{G}^{\mu\nu} = i \epsilon^{\mu\nu\alpha\beta} q_{\alpha}
\left[ p_\beta ( \pslash \gamma_5 \widehat G_p
	       + \qslash \gamma_5 \widehat G_q )
     + \gamma_\beta \gamma_5 \widehat G_\gamma 
\right].
\ene
Once again, one finds that three functions, $\widehat G_i$, can be
constructed in terms of scalar and pseudo-scalar vertices.
However, in this case there is a new feature, {\em the function
$\widehat G_p$ does not contribute for a free nucleon}, whereas it
does contribute in a nucleus.
\begin{figure}[hbt]
\begin{center}
\epsfig{file=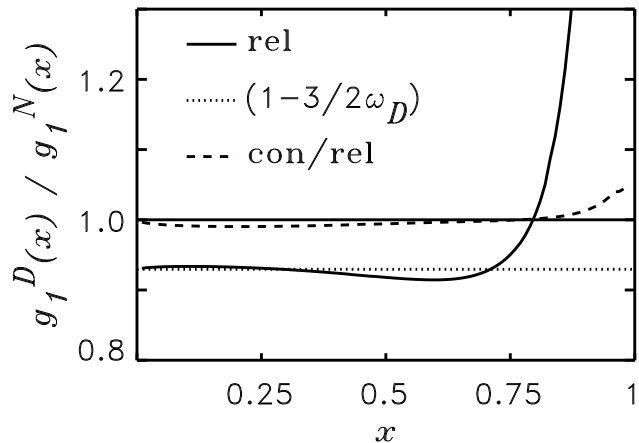,height=7cm}
\caption{Ratio of deuteron and nucleon structure functions in the
	full model (solid), and with a constant depolarization factor
	corresponding to $\omega_D = 4.7\%$ (dotted line).
	The dashed curve is the ratio of $g_1^D$ calculated via
	convolution to $g_1^D$ calculated in the relativistic model
	--- from Ref.\protect\cite{MPT}.}
\end{center}
\end{figure}

The spin-dependent deuteron structure function is given by the trace of
$\widehat{G}^{\mu\nu}$ with a spin-dependent $ND$ amplitude \cite{MPT},
which can be evaluated using the relativistic $DNN$ vertex of Ref.\cite{BG}.
Surprisingly, Fig.8 shows that the ratio of the convolution approximation
to the fully off-shell calculation (the dashed curve) is even closer to
unity in this case than in the spin-independent case.

The comparison between the solid and dotted curves in Fig.8 shows that
Eq.(\ref{eq:3.3}) is reliable at the 2\% level for $x$ below 0.7.
However, the excellent agreement in this region between (\ref{eq:3.3})
and the full calculation relies on a knowledge of the deuteron $D$-state
probability.
As shown in Ref.\cite{MPT}, a change of $\omega_D$ by 2\% (e.g. from 4\%
to 6\%) leads to an error of order 10\% or more in $g_1^n$.
For $x > 0.7$, on the other hand, the approximation (\ref{eq:3.3})
fails dramatically.
This will be extremely important when testing the predictions of PQCD
for the $x \ra 1$ behavior of the polarized distributions in
Eqs.(\ref{Deltau})--(\ref{Deltad}).

\section{Conclusions}

As we have seen, there is now overwhelming experimental evidence for a
large, non-perturbative component of the non-strange sea of the nucleon.
This is almost certainly associated with the pion cloud of the nucleon.
A full, quantitative analysis of the relevant data, especially the
Drell-Yan data from FNAL, will provide important new insight into the
process of dynamical symmetry breaking in QCD. 

Having seen the importance of the non-perturbative component of the
non-strange sea, it is natural to ask about the strange and even the
charm sea. For the former there is, as yet, no evidence for an asymmetry
between $s$ and $\overline s$ quarks -- although there are limits on how
big it could be. On the other hand, in the case of charm there is
tremendous interest in a possible non-perturbative component -- the
intrinsic charm sea. Amongst other things this is important for the
interpretation of the anomaly at high invariant mass observed recently
at HERA. From the theoretical point of view it is a totally open
question whether or not there is an intrinsic charm sea and, if so,
whether or not it is asymmetric.

The large-$x$ region of the parton distributions corresponds to the high
momentum components of the nucleon wave function. It is a vital question
for our understanding of hadron structure just how these high momentum
components are generated. We have seen that the most recent analysis of
the structure function of the deuteron strongly supports the suggestion
that these high momentum components are generated by gluon final state
interactions which can be understood in terms of perturbative QCD. As a
consequence the valence $d/u$ ratio seems not to vanish as $x \ra 1$.
This idea needs further testing but certainly has important consequences
for event rates at machines like HERA.

Our new understanding of the unpolarized parton distributions at
large-$x$ also leads us to new expectations for the polarized
distributions -- especially for the polarized neutron. The extraction of
this information using a bound neutron target requires a sophisticated
understanding of off-shell corrections in nuclear deep-inelastic
scattering. 

\section*{Acknowledgments}
It is a pleasure to acknowledge the contributions to the work described
here by W. Melnitchouk, G. Piller and A. Schreiber. I would also like to
thank Prof. Dong-Pil Min and the other staff of the Asia Pacific Center
for Theoretical Physics for their hospitality during this workshop, held
in honour of Mannque Rho's 60th birthday. This work was
supported by the Australian Research Council.

\section*{References}

\begin{thebibliography}{99}
%
\bibitem{LL} F. X. Lee and D. B. Leinweber, hep-lat/9711044.
%
\bibitem{DISS} A. W. Thomas,
Prog. Part. Nucl. Phys. {\bf 20} (1988) 21.
%
%
\bibitem{E866}
E.A. Hawker {\em et al.} (E866 Collaboration),
``Measuring the $\bar u-\bar d$ Asymmetry in the Proton Sea'',
presented at XXXII Moriond Conference, 22-29 March 1997.
%
\bibitem{MT_fewb} A. W. Thomas and W. Melnitchouk, hep-ph/9708484.
%
\bibitem{IM}
T. Inoue and T. Maskawa,
Prog. Theor. Phys. {\bf 54} (1975) 1833.
%
\bibitem{CHT}
A. Chodos and C.B. Thorn,
Phys. Rev. D {\bf 12} (1975) 359.
%
\bibitem{BRR}
G.E. Brown and M. Rho,
Phys. Lett. B {\bf 82} (1979) 177.
%
\bibitem{CBM}
S. Th\'eberge, G.A. Miller and A.W. Thomas,
Phys. Rev. D {\bf 22} (1980) 2838;
{\em ibid} D {\bf 23} (1981) 2106(e);
%
A.W. Thomas,
Adv. Nucl. Phys. {\bf 13} (1984) 1;
%
G.A. Miller,
Int. Rev. Nucl. Phys. {\bf 2} (1984) 190.
%
\bibitem{SULL}
J.D. Sullivan,
Phys. Rev. D {\bf 5} (1972) 1732.
%
\bibitem{CLLS}
C.H. Llewellyn Smith,
Phys. Lett. B {\bf 128} (1983) 107.
%
\bibitem{ET}
M. Ericson and A.W. Thomas,
Phys. Lett. B {\bf 128} (1983) 112.
%
\bibitem{THO83}
A.W. Thomas,
Phys. Lett. B {\bf 126} (1983) 97.
%
\bibitem{FMS}
L.L. Frankfurt, L. Mankiewicz and M.I. Strikman,
Zeit. Phys. A {\bf 334} (1989) 334;
%
W. Koepf, L.L. Frankfurt and M.I. Strikman,
Phys. Rev. D {\bf 53} (1996) 2586.
%
\bibitem{NMCO}
P. Amaudruz {\it et al.} (NMC),
Phys. Lett. B {\bf 295} (1992) 159;
Phys. Rev. Lett. {\bf 66} (1991)~2712.
%
\bibitem{SPTH}
J. Speth and A.W. Thomas,
``Mesonic Contributions to the Spin and Flavor Structure of the
Nucleon''
(J\"ul-3283, Sept. 1996), to appear in Adv. Nucl. Phys. (1998).
%
\bibitem{HM}
E.M. Henley and G.A. Miller,
Phys. Lett. B {\bf 251} (1990) 453.
%
\bibitem{SST}
A.I. Signal, A.W. Schreiber and A.W. Thomas,
Mod. Phys. Lett. A {\bf 6} (1991) 271.
%
\bibitem{KL}
S. Kumano and J.T. Londergan,
Phys. Rev. D {\bf 44} (1991) 717;
S. Kumano, hep-ph/9702367.
%
\bibitem{HSB}
W.-Y.P. Hwang, J. Speth and G.E. Brown,
Zeit. Phys. A {\bf 339} (1991) 383.
%
\bibitem{MTLC}
W. Melnitchouk and A.W. Thomas,
Phys. Rev. D {\bf 47} (1993) 3794.
%
\bibitem{ZOLL}
V.R. Zoller,
Z. Phys. C {\bf 60} (1993) 141.
%
\bibitem{ST}
A.I. Signal and A.W. Thomas,
Phys. Lett. B {\bf 191} (1987) 205.
%
\bibitem{BROD}
S.J. Brodsky and B.-Q. Ma,
Phys. Lett. B {\bf 381} (1996) 317.
%
\bibitem{SMSBAR}
X. Ji and J. Tang,
Phys. Lett. B {\bf 362} (1995) 182;
%
H. Holtmann {\em et al.},
Nucl. Phys. A {\bf 569} (1996) 631;
%
W. Melnitchouk and M. Malheiro,
Phys. Rev. C {\bf 55} (1997) 431.
%
\bibitem{CCFR}
A. Bazarko {\em et al.} (CCFR Collaboration),
Zeit. Phys. C {\bf 65} (1995) 189.
%
\bibitem{GV}
J.F. Gunion and R. Vogt,
UCD-97-14, LBNL-40399 [hep-ph/9706252].
%
\bibitem{BHPS}
S.J. Brodsky, P. Hoyer, C. Peterson and N. Sakai,
Phys. Lett. {\bf 93} B, 451 (1980);
%
S.J. Brodsky, C. Peterson and N. Sakai,
Phys. Rev. D {\bf 23}, 2745 (1981).
%
\bibitem{NNNT}
F.S. Navarra, M. Nielsen, C.A.A. Nunes and M. Teixeira,
Phys. Rev. D {\bf 54}, 842 (1996).
%
\bibitem{PNNDB}
S. Paiva, M. Nielsen, F.S. Navarra, F.O. Duraes and L.L. Barz,
IFUSP-P-1240 [hep-ph/9610310].
%
\bibitem{KLT}
S. Kuhlmann, H.L. Lai and W.K. Tung,
Phys. Lett. B {\bf 409} (1997) 271.
%
\bibitem{MT_HERA}
W. Melnitchouk and A.W. Thomas,
hep-ph/9707387.
%
\bibitem{CLO79}
F.E. Close,
{\em An Introduction to Quarks and Partons}
(Academic Press, 1979).
%
\bibitem{CLO73}
F.E. Close,
Phys. Lett. B {\bf 43} (1973) 422.
%
\bibitem{CAR75}
R. Carlitz,
Phys. Lett. B {\bf 58} (1975) 345.
%
\bibitem{CT}
F.E. Close and A.W. Thomas,
Phys. Lett. B {\bf 212} (1988) 227.
%
\bibitem{FJ}
G.R. Farrar and D.R. Jackson,
Phys. Rev. Lett. {\bf 35} (1975) 1416.
%
\bibitem{BBS}
S.J. Brodsky, M. Burkardt and I. Schmidt,
Nucl. Phys. {\bf B441} (1995) 197.
%
\bibitem{WHIT}
L.W. Whitlow {\em et al.},
Phys. Lett. B {\bf 282} (1992) 475.
%
\bibitem{GOMEZ}
J. Gomez {\em et al.},
Phys. Rev. D {\bf 49} (1994) 4348.
%
\bibitem{FS}
L.L. Frankfurt and M.I. Strikman,
Phys. Lett. B {\bf 76}  (1978) 333;
Phys. Rep. {\bf 76} (1981) 215.
%
\bibitem{KU}
L.P. Kaptari and A.Yu. Umnikov,
Phys. Lett. B {\bf 259} (1991) 155.
%
\bibitem{LG}
F. Gross and S. Liuti,
Phys. Rev. C {\bf 45} (1992) 1374;
%
\bibitem{BT}
M.A. Braun and M.V. Tokarev,
Phys. Lett. B {\bf 320} (1994) 381.
%
\bibitem{DT}
G.V. Dunne and A.W. Thomas,
Nucl. Phys. A {\bf 455} (1986) 701.
%
\bibitem{NW}
K. Nakano and S.S.M. Wong,
Nucl. Phys. A {\bf 530} (1991) 555.
%
\bibitem{FS88}
L.L. Frankfurt and M.I. Strikman,
Phys. Rep. {\bf 160} (1988) 235.
%
\bibitem{MILL}
H. Jung and G.A. Miller,
Phys. Lett. B {\bf 200} (1988) 351.
%
\bibitem{CONV}
R.L. Jaffe,
in {\em Relativistic Dynamics and
Quark-Nuclear Physics},
eds. M.B.Johnson and A.Pickleseimer
(Wiley, New York, 1985).
S.V. Akulinichev, S.A. Kulagin and G.M. Vagradov,
Phys. Lett. B {\bf 158} (1985) 485;
S.A. Kulagin, G. Piller and W. Weise,
Phys. Rev. C {\bf 50} (1994) 1154.
%
\bibitem{MST1}
W. Melnitchouk, A.W. Schreiber and A.W. Thomas,
Phys. Rev. D {\bf 49} (1994) 1183.
%
\bibitem{MSM}
P. Mulders, A.W. Schreiber and H. Meyer,
Nucl. Phys. A {\bf 549} (1992) 498.
%
\bibitem{MST2}
W. Melnitchouk, A.W. Schreiber and A.W. Thomas,
Phys. Lett. B {\bf 335} (1994) 11.
%
\bibitem{EMC}
European Muon Collaboration, J.J. Aubert {\em et al.},
Phys. Lett. B {\bf 110} (1982) 73; 
Nucl. Phys. B {\bf 213} (1983) 213.
%
\bibitem{US}
T. Uchiyama and K. Saito,
Phys. Rev. C {\bf 38} (1988) 2245.  
%
\bibitem{LIGHTFRONT}
J. Carbonell, B. Desplanques, V.A. Karmanov and J.-F.Mathiot,
to appear in Phys.~Rep.
%
\bibitem{MTNP}
W. Melnitchouk and A.W. Thomas,
Phys. Lett. B {\bf 377} (1996) 11.
%
\bibitem{BODEK}
A. Bodek {\em et al.},
Phys. Rev. D {\bf 20} (1979) 1471;
A. Bodek and J.L. Ritchie,
Phys. Rev. D {\bf 23} (1981) 1070.
%
\bibitem{CDHS}
H. Abramowicz {\em et al.} (CDHS Collaboration),
Z. Phys. C {\bf 25} (1983) 29.
%
\bibitem{MST2}
W. Melnitchouk, A.W. Schreiber and A.W. Thomas,
Phys. Lett. B {\bf 335} (1994) 11.
%
\bibitem{H1}
C. Adloff {\em et al.} (H1 Collaboration),
Zeit. Phys. C {\bf 74} (1997) 191.
%
\bibitem{ZEUS}
J. Breitwig {\em et al.} (ZEUS Collaboration),
Zeit. Phys. C {\bf 74} (1997) 207.
%
\bibitem{CHENG}
H.-Y. Cheng,
Int. J. Mod. Phys. A {\bf 11} (1996) 5109;
M. Anselmino {\em et al.},
Phys. Rep. {\bf 261} (1995) 1.
%
\bibitem{CTEQ}
H.L. Lai, J. Huston, S. Kuhlmann, F. Olness, J.F. Owens,
D. Soper, W.K. Tung and H. Weerts,
Phys. Rev. D {\bf 55}, 1280 (1997).
%
\bibitem{H196}
H1 Collaboration, S. Aid {\em et al.},
Phys. Lett. B {\bf 379}, 319 (1996).
%
\bibitem{BABU}
K.S. Babu, C. Kolda and J. March-Russell,
Phys. Lett. B {\bf 408} (1997) 268.
%
\bibitem{EMC_SPIN}
J. Ashman {\em et al.} (EMC),
Phys. Lett. B {\bf 206} (1988) 364.
%
\bibitem{VER}
S.A. Larin, T. van Ritbergen and J.A.M. Vermaseren,
Phys. Lett. B {\bf 404} (1997) 153.
%
\bibitem{DPOL}
D. Adams {\em et al.} (SMC),
Phys. Lett. B {\bf 357} (1995) 248;
K. Abe {\em et al.} (E143 Collaboration),
Phys. Rev. Lett. {\bf 75} (1995) 25.
%
\bibitem{MPT}
W. Melnitchouk, G. Piller and A.W. Thomas,
Phys. Lett. B {\bf 346} (1995) 165;
G. Piller, W. Melnitchouk and A.W. Thomas,
Phys. Rev. C {\bf 54} (1996) 894.
%
\bibitem{KMPW}
S.A. Kulagin, W. Melnitchouk, G. Piller and W. Weise,
Phys. Rev. C {\bf 52} (1995) 932.
%
\bibitem{BG}
W.W. Buck and F. Gross,
Phys. Rev. D {\bf 20} (1979) 2361;
R.G. Arnold, C.E. Carlson and F. Gross,
Phys. Rev. C {\bf 21} (1980) 1426;
F. Gross, J.W. Van Orden and K. Holinde,
Phys. Rev. C {\bf 45} (1992) 2094.
%
\end{thebibliography}
\end{document}